# AN ALTERNATIVE APPROACH TO THE PHONON THEORY OF LIQUIDS: EVOLUTION OF THE ENERGY OF DIFFUSION

M. Y. Esmer[1, a)] and Bahtiyar A. Mamedov[1]

[1]*Department of Physics, Faculty of Arts and Sciences, Gaziosmanpaşa University, Tokat, Türkiye*
[a)] Authors to whom correspondence should be addressed: mylmzesmer@hotmail.com

## ABSTRACT

With regard to the three basic states of matter (solid, liquid, gas), the calculation of the heat capacity of liquids in a general form has been considered one of the deepest and most interesting challenges in condensed matter physics, due to the strong, system-specific interactions involved. Notwithstanding the theoretical difficulties, there have recently been significant advances in our understanding of liquids, and the phonon theory of liquids has been proposed. However, this theory uses the virial theorem to calculate the liquid energy. Here, we propose an alternative version of the phonon theory of liquids by taking into account the numbers of oscillating atoms and diffusing atoms, rather than using the virial theorem. A new formula is derived for the liquid energy in both the quantum and the classical anharmonic cases. To verify the proposed approach, theoretical predictions are compared with the experimental specific heat of liquid mercury. Finally, we obtain a new expression for the energy of a supercritical system.

## I. INTRODUCTION

Studies reveal that as temperature increases, the heat capacity at constant volume typically declines for simple liquids, from around $3k_B$ per atom at the melting point to about $2k_B$ at higher temperatures [1, 2]. This behavior cannot be explained in the same straightforward manner as for the solid and gas phases, because the interactions in a liquid are strongly system-dependent. It has therefore been asserted that a universal way of calculating the energy of a liquid is not feasible [3, 25].

Historically, the study of liquids has mainly been approached from the perspective of the gas phase, with a focus on methods of calculating both the interaction energy and the kinetic energy

[4]. This approach requires detailed information about interatomic interactions and correlation functions, which are generally complex, with no universal method of deriving them. In addition, the decrease in liquid heat capacity observed experimentally cannot be sufficiently explained with this method [5, 6].

Frenkel proposed an alternative approach in which the problem was viewed from the perspective of the solid phase; he introduced the concept of the liquid relaxation time as the average time between particle jumps at one point in space, and predicted that liquids have a solid-like ability to support the shear mode with frequency larger than $1/\tau$ [7]. However, the authorities of the time did not support this perspective, and it was eventually forgotten [25].

After a gap of roughly 50 years, Frankel's prediction that liquids could support a solid-like frequency was experimentally verified [8-10]. In recent years, Trachenko and his co-workers have developed alternative approaches to modeling the heat capacity of liquids that incorporate Frankel's concept and the supporting experimental findings [11-13]. This model, which is known as the phonon theory of liquids, was tested by researchers based on the heat capacities of more than 20 different systems, and strong agreement was found between the experimental data and the theoretical predictions [13]. An independent, thorough verification of the phonon theory of liquids was also conducted by Proctor [14,15]. This theory has been widely adopted as a theoretical framework for both applied and fundamental research in various fields from nanofluidics to planetary science [16-20].

However, this theory utilizes the virial theorem (or the equipartition theorem) to calculate the liquid energies [11-13], causing anharmonic effects to be included in the energy of diffusion. In this work, we present a new way to calculate the energy of diffusion in liquids on the basis of the numbers of oscillating atoms and diffusing atoms, rather than focusing on the virial theorem. A new analytical formula is developed to describe liquid energy in both quantum and classical anharmonic scenarios, and the validity of this formula is tested by comparing theoretical predictions to experimental data on the specific heat of liquid mercury. In addition, we apply our method to a supercritical system, and propose a new expression for the energy of a supercritical system as an alternative to the phonon theory of liquids. Finally, we discuss some advantages of this method.

## II. ENERGY OF A SUBCRITICAL LIQUID

In this section, we start by reviewing the phonon theory of liquid thermodynamics as explained in Refs. [11-13], which was originally proposed by Trachenko for classical harmonic liquids. According to this theory, a liquid exhibits two types of atomic motion: the first is vibrational motion, which includes one longitudinal mode and two shear modes with frequency $\omega > \omega_F$, where $\omega_F$ is the Frenkel frequency; and the second is diffusive jump motion between two equilibrium positions. The liquid energy can be expressed as

$$E = K_l + P_l + K_s(\omega > \omega_F) + P_s(\omega > \omega_F) + K_d + P_d \qquad (1)$$

$$E = K + P_l + P_s(\omega > \omega_F) + P_d \qquad (2)$$

where $K$ is the total kinetic energy of the liquid; $K_l$ and $P_l$ are the kinetic and potential parts of the longitudinal phonon energy; $K_s(\omega > \omega_F)$ and $P_s(\omega > \omega_F)$ are the transverse kinetic and potential energies with frequency $\omega > \omega_F$; and $K_d$ and $P_d$ are the kinetic and potential parts of the energy of the diffusing atoms, respectively [11-13]. Using the equipartition theorem (or the virial theorem), which implies $K = \frac{E_l + E_s}{2}$, $P_l = \frac{E_l}{2}$, and $P_s(\omega > \omega_F) = \frac{E_s(\omega > \omega_F)}{2}$, Eq. (2) was re-written by Trachenko as

$$E = E_l + E_s(\omega > \omega_F) + \frac{E_s(\omega < \omega_F)}{2} \qquad (3)$$

and each term in Eq. (3) was calculated with an approach similar to that used in the harmonic theory of solids [11-13]. Then, the total liquid energy can be obtained for the classical case as

$$E = NT\left(1 + \frac{\alpha T}{2}\right)\left(3 - \left(\frac{\omega_F}{\omega_D}\right)^3\right) \qquad (4)$$

[12] and for the quantum case as

$$E = NT\left(1 + \frac{\alpha T}{2}\right)\left(3D\left(\frac{\hbar \omega_D}{T}\right) - \left(\frac{\omega_F}{\omega_D}\right)^3 D\left(\frac{\hbar \omega_F}{T}\right)\right) \qquad (5)$$

[13], where $\alpha$ is the coefficient of thermal expansion and takes into account the effect of thermal expansion ($k_B = 1$), $\omega_D$ is the Debye frequency, and $D(x)$ is the Debye function [21].

From our point of view, to establish a relation between the energy of vibrational motion and the energy of diffusional motion, the authors of the studies [11-13] perform two crucial steps in the calculation of the liquid energy. In the first, the total kinetic energy of the system is given as $K = K_l + K_s + K_d$, meaning that it is independent of how the kinetic energy is divided between the oscillating and diffusive components, while in the second, the virial theorem is used to provide a relationship between the total kinetic energy of the system and the energies of longitudinal and transverse modes, $K = \frac{E_l + E_s}{2}$.

The authors of studies [7, 11–13] have already determined the type of motion of particles in liquids as vibrational and diffusional motion; however, it is still necessary to determine how many particles exhibit diffusional motion and how many exhibit vibrational motion. This is because as the proportion of particles undergoing vibrational motion increases, the liquid tends to exhibit properties closer to that of a solid, whereas an increase in the proportion of particles engaged in diffusion causes the liquid to behave more similarly to a gas. The type of motion of particle in a liquid can be controlled by temperature and pressure.

Since the motion of a particle in a liquid consists of vibrational and diffusional parts, the total liquid energy can be written as

$$E = K_l + P_l + K_s(\omega > \omega_F) + P_s(\omega > \omega_F) + K_d + P_d \tag{6}$$

As explained above, Trachenko calculated the expressions for potential and kinetic energies using the equipartition theorem (or the virial theorem). This step is not necessary in our approach, and the total liquid energy can be re-written as

$$E = E_l + E_s(\omega > \omega_F) + E_d \tag{7}$$

using the energy of the longitudinal mode $E_l = K_l + P_l$, the energy of the shear modes with frequency $\omega > \omega_F$ $E_s = K_s(\omega > \omega_F) + P_s(\omega > \omega_F)$, and the energy of diffusion $E_d = K_d + P_d$. We can establish a relationship between the energy of vibration and the energy of diffusion based on the number of particles (the number of atoms), as the number of the atoms does not change during the phase transition. Since all atoms vibrate, there are approximately $3N$ modes in a liquid around the melting point, where $N$ is the number of particles in the system. As the temperature increases, the number of diffusing atoms increases, and the number of oscillating particles decreases, meaning that the number of modes begins to decrease. We can approximately determine the numbers of oscillating particles and diffusing particles using the following calculations. A liquid supports one longitudinal mode and two shear modes with frequency larger than $\omega_F$. There are then $N$ longitudinal modes in the system, and the number of shear modes with frequency $\omega > \omega_F$ can be calculated from the density of states in the Debye theory as

$$N_s = \int_{\omega_F}^{\omega_D} g_s(\omega) d\omega = \int_{\omega_F}^{\omega_D} 6N \frac{\omega^2}{\omega_D^3} d\omega = 2N\left(1 - \left(\frac{\omega_F}{\omega_D}\right)^3\right) \tag{8}$$

where $g_s(\omega) = 6N\frac{\omega^2}{\omega_D^3}$ is the density of states for the shear modes. The total number of one longitudinal and two shear modes with frequency $\omega > \omega_F$ then becomes $N + 2N\left(1 - \left(\frac{\omega_F}{\omega_D}\right)^3\right)$. If an oscillating atom in a liquid has three degrees of freedom, the system has $3N_v$ modes, where $N_v$ is the number of oscillating atoms. Using these two expressions, we can obtain the number of oscillating atoms as

$$3N_v = N + 2N\left(1 - \left(\frac{\omega_F}{\omega_D}\right)^3\right) \tag{9}$$

$$N_v = N - \frac{2N}{3}\left(\frac{\omega_F}{\omega_D}\right)^3 \tag{10}$$

Note that the number of oscillating atoms is smaller than the number of atoms in the liquid. We can finally calculate the number of diffusing atoms in the liquid as

$$N_d = \frac{2N}{3}\left(\frac{\omega_F}{\omega_D}\right)^3 \tag{11}$$

using $N_v + N_d = N$. This expression enables us to calculate the energy of diffusion in Eq. (7). According to Ref. [13], the potential parts of the energy of diffusing atoms, $P_d$, can be neglected since $P_d \ll P_s(\omega > \omega_F)$; in this case, similarly to an ideal gas, the total energy of the diffusing atoms consists only of kinetic energy. Using the total energy of an ideal gas of $N_d$ particles, we finally obtain the energy of diffusion:

$$E_d = \frac{3}{2}N_d T = N\left(\frac{\omega_F}{\omega_D}\right)^3 T. \tag{12}$$

The first two terms in Eq. (7) were already calculated as $E_l + E_s(\omega > \omega_F) = N\left(3 - 2\left(\frac{\omega_F}{\omega_D}\right)^3\right)T$ by Trachenko [11]. By combining this expression and Eq. (12) with Eq. (7), we obtain the following relation for the liquid energy:

$$E = N\left(3 - 2\left(\frac{\omega_F}{\omega_D}\right)^3\right)T + N\left(\frac{\omega_F}{\omega_D}\right)^3 T = N\left(3 - \left(\frac{\omega_F}{\omega_D}\right)^3\right)T. \qquad (13)$$

This equation is the same as that derived by Trachenko in the harmonic approximation [11], but is slightly different in the anharmonic case.

In our analysis, we assume that the phonon frequency is independent of temperature. However, since thermal expansion and anharmonic effects play a significant role in liquids, the phonon frequencies decrease with temperature, and we therefore need to recalculate the first two terms in Eq. (7) corresponding to the vibrational motion, taking into account anharmonicity and thermal expansion. These terms have been calculated by Bolmatov and Trachenko [12] as

$$E_l + E_s(\omega > \omega_F) = \left(1 + \frac{\alpha T}{2}\right)\left(3 - 2\left(\frac{\omega_F}{\omega_D}\right)^3\right)NT. \qquad (14)$$

Since the atoms that undergo diffusive motion do not perform vibrational motion at the same time, we do not include the anharmonic effect in the diffusional part of the energy. Adding Eq. (14) to the energy of diffusion calculated above, $N\left(\frac{\omega_F}{\omega_D}\right)^3 T$, gives the anharmonic liquid energy:

$$E = \left(1 + \frac{\alpha T}{2}\right)N\left(3 - 2\left(\frac{\omega_F}{\omega_D}\right)^3\right)T + N\left(\frac{\omega_F}{\omega_D}\right)^3 T \qquad (15)$$

In the anharmonic case, our equation for the liquid energy is slightly different from Eq. (4) obtained by Bolmatov and Trachenko. However, at low temperature, where $\omega_D \gg \omega_F$, Eq. (15) gives the same result as the findings of Bolmatov and Trachenko:

$$E = 3NT\left(1 + \frac{\alpha T}{2}\right). \qquad (16)$$

Our approach is also applicable to the quantum regime. In this case, we need to recalculate the first two energy terms corresponding to the vibrational motion in Eq. (7), taking into account the quantum regime. In study [13], the authors computed these terms based on the phonon theory of liquids, providing the following expressions:

$$E_l + E_s(\omega > \omega_F) = NT\left(1 + \frac{\alpha T}{2}\right)\left(3D\left(\frac{\hbar\omega_D}{T}\right) - 2\left(\frac{\omega_F}{\omega_D}\right)^3 D\left(\frac{\hbar\omega_F}{T}\right)\right). \qquad (17)$$

Substituting Eq. (17) and the energy of diffusion calculated above into Eq. (7) gives the liquid energy in the quantum regime:

$$E = NT\left(1 + \frac{\alpha T}{2}\right)\left(3D\left(\frac{\hbar\omega_D}{T}\right) - 2\left(\frac{\omega_F}{\omega_D}\right)^3 D\left(\frac{\hbar\omega_F}{T}\right)\right) + N\left(\frac{\omega_F}{\omega_D}\right)^3 T. \qquad (18)$$

Note that Eqs. (15) and (18) explain the decrease in liquid heat capacity from about 3 to 2. In the next section, we compare the results calculated using Eqs. (15) and (18) with experimental heat capacity data.

### III. EXPERIMENTAL DATA COMPARISON

In the previous section, we obtained formulae for the energy of liquids in both the quantum and classical cases, as given in Eqs. (15) and (18), using a new alternative approach. In this section, we compare them with values for the experimental heat capacity and those given in the original work to test our predictions.

Eqs. (15) and (18) depend on the experimental parameter $\omega_F$. By applying the Maxwell relationship $\tau = \frac{\eta}{G_\infty}$, where $G_\infty$ is the infinite-frequency shear modulus and $\eta$ is viscosity, we can formulate $\omega_F = \frac{2\pi}{\tau} = \frac{2\pi G_\infty}{\eta}$ [11-13]. To ascertain the value of $\eta$, the Vogel-Fulcher-Tammann (VFT) model, expressed as $\eta = \eta_0 \exp\left(\frac{A}{T-T_0}\right)$, is frequently applied. This is done by fitting the VFT equation to experimental viscosity data.

For simplicity, we have used mercury, a simple monatomic liquid. The experimental viscosity data for $Hg$ has been taken from Ref. [22] and fitted using the VFT model, expressed as $\eta = \eta_0 \exp\left(\frac{A}{T-T_0}\right)$. The fitting parameters used here are $\eta_0$, $A$ and $T_0$, with values drawn from Ref. [11]. The analytical expressions for the mercury heat capacity per atom are calculated by using the definition $c_v = \frac{1}{N}\frac{dE}{dT}$, Eqs. (15) and (18), $\omega_F = \frac{2\pi G_\infty}{\eta}$ and the VFT model. The calculated results for the heat capacity of liquid mercury as a function of temperature are presented in Figs. 1 and 2.

Fig. 1 presents a comparison of the heat capacity for liquid mercury calculated from anharmonic theory, using Eq. (15), with the work of Bolmatov and Trachenko [12] and the experimental heat capacity drawn from Ref. [1]. In Fig. 1, the solid green curve represents the anharmonic heat capacity for liquid mercury calculated by Bolmatov and Trachenko using Eq. (4), and the blue curve shows the results from our method calculated using Eq. (15). Similarly, in Fig. 2 we show a comparison of the heat capacities for liquid mercury in the quantum case with experimental data. As can be seen from Figs. 1 and 2, we find good agreement between the calculated and experimental heat capacities per atom. This is not surprising, since our approach is based on the phonon theory of liquid, and explains the decrease in heat capacity from 3 to 2, in a similar way to the phonon theory of liquids.

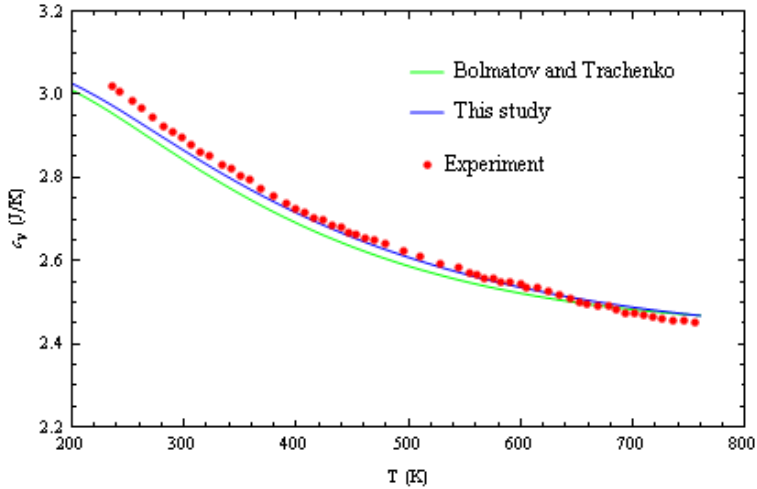

FIG. 1. Comparison of theoretical predictions in the anharmonic case with experimental data on the heat capacity of mercury. The red points represent experimental data on the heat capacity of liquid mercury [1]. The blue curve (top) shows the results from this study, calculated using Eq. (15) in the anharmonic case, where $\alpha = 1.830 \times 10^{-4}$ K$^{-1}$ and $\tau_D G_\infty = 0.618 \times 10^{-3}$ Pa s. The green curve (bottom) shows the results from the work of Bolmatov and Trachenko, calculated using Eq. (4) in the anharmonic case, where $\alpha = 1.60 \times 10^{-4}$ K$^{-1}$ and $\tau_D G_\infty = 6.272 \times 10^{-4}$ Pa s [12]. The experimental value of $\alpha$ for mercury is $1.8 \times 10^{-4}$ K$^{-1}$ [13].

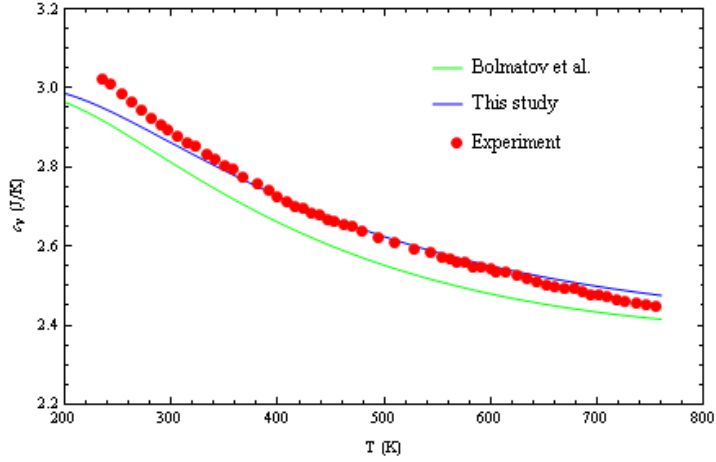

FIG. 2. Comparison of theoretical predictions in the quantum case with experimental data on the heat capacity of mercury. The red points represent experimental data on the heat capacity of liquid mercury [1]. The blue curve (top) shows the results from this study, calculated using Eq. (18) in the quantum case, where $\alpha = 1.83 \times 10^{-4}$ K$^{-1}$, $G_\infty = 1.271$ GPa and $\tau_D = 0.49$ ps. The green curve (bottom) represents the results of work by Bolmatov et al., calculated using Eq. (5) in the quantum case, where $\alpha = 1.60 \times 10^{-4}$ K$^{-1}$, $G_\infty = 1.31$ GPa and $\tau_D = 0.49$ ps [13]. The experimental value of $\alpha$ for mercury is $1.8 \times 10^{-4}$ K$^{-1}$ [13].

## IV.  ENERGY OF A SUPERCRITICAL SYSTEM

In Sec. II, we showed that our approach explains the decrease in $c_v$ from 3 to 2. As the temperature increases, $c_v$ is expected to drop below 2 and to become closer to the ideal-gas value of 3/2. This decrease has recently been clarified by a new dynamic line on a phase diagram, the Frenkel line (FL) [23-25,16]. Below the FL, particles exhibit both oscillatory and diffusive behavior, allowing the system to support rigidity and high-frequency transverse modes; above the FL, particle motion is solely diffusive, resulting in the loss of rigidity and the inability to support transverse modes across all available frequencies [23-25,16]. On the basis of this picture, using the equipartition theorem, the energy of non-rigid gas-like supercritical fluid above the FL can be calculated as:

$$E = \frac{3}{2}NT + \frac{1}{2}NT\left(1 + \frac{\alpha T}{2}\right)\left(\frac{a}{L}\right)^3, \qquad (19)$$

where $a$ is the interatomic separation, and $L$ is the mean free path of particles [16]. When $L \approx a$ at the Frenkel line, Eq. (19) gives $c_v = 2$, whereas when $L \gg a$ at high temperature, it gives $c_v = $

3/2, the expected ideal-gas value. In this section, the novel aspect of this work is that we propose our own version of this equation that explains the decrease in $c_v$ from 2 to 3/2.

We now turn to recalculating the energy of a supercritical system using our model. Above the FL, two transverse modes vanish, and the oscillations of the longitudinal mode decrease with temperature, starting with the highest frequency. The liquid energy, Eq. (7), can then be rewritten for the energy of the supercritical system as

$$E = E_l(\omega < \omega_L) + E_D + NT \tag{20}$$

where $\omega_L$ is the minimal frequency, $E_l(\omega < \omega_L)$ is the energy of longitudinal mode with $\omega < \omega_L$, and $E_D$ is the energy of the diffusing atoms due to the disappearance of the remaining longitudinal mode. When $\omega_F = \omega_D$, the two transverse modes in Eq. (7) are completely lost at all frequencies, $E_s(\omega > \omega_F) = 0$, and $E_d = NT$. The first term in Eq. (20) is given as $E_l(\omega < \omega_L) = NT\left(1 + \frac{\alpha T}{2}\right)\left(\frac{\omega_L}{\omega_D}\right)^3 = NT\left(1 + \frac{\alpha T}{2}\right)\left(\frac{a}{L}\right)^3$ in Ref. [25]. We can use our method given above to calculate $E_D$ in Eq. (20) rather than using the equipartition theorem, as follows.

At the FL, according to Eq. (11), $\omega_F = \omega_D$ gives $N_d = 2N/3$. Thus, $N/3$ atoms, corresponding to $N$ longitudinal modes, undergo oscillatory motion. Above the FL, the system enters the new dynamic regime and the longitudinal mode becomes *gapped*. The number of longitudinal modes with frequency $\omega < \omega_L$ can then be calculated in the same way as mentioned in Sec. II, as $N_L = \int_0^{\omega_L} 3N \frac{\omega^2}{\omega_D^3} d\omega = N\left(\frac{\omega_L}{\omega_D}\right)^3$, corresponding to $N_V = \frac{N}{3}\left(\frac{\omega_L}{\omega_D}\right)^3$ oscillating atoms since the system has $3N_V$ modes. We can finally calculate the number of diffusing atoms, $N_D$, due to the loss of longitudinal modes as

$$N_D = \frac{N}{3}\left(1 - \left(\frac{\omega_L}{\omega_D}\right)^3\right) \tag{21}$$

using $N_V + N_D = N/3$. From the total energy of an ideal gas of $N_D$ particles, we finally obtain the energy of diffusion:

$$E_D = \frac{NT}{2}\left(1 - \left(\frac{\omega_L}{\omega_D}\right)^3\right) \tag{22}$$

From all terms in Eq. (20) and using the expression $\left(\frac{\omega_L}{\omega_D}\right)^3 = \left(\frac{a}{L}\right)^3$ in Ref. [16], we finally obtain the energy of the supercritical system as

$$E = \frac{3NT}{2} + NT\left(1 + \frac{\alpha T}{2}\right)\left(\frac{a}{L}\right)^3 - \frac{NT}{2}\left(\frac{a}{L}\right)^3 \qquad (23)$$

Note that Eqs. (19) and (23) are the same for $\alpha = 0$, whereas when $L \gg a$ at high temperature, Eq. (23) gives $c_v = 3/2$, the ideal gas value as expected. When $L \approx a$ at the Frenkel line, this approach gives the same result as in Eq. (19), $c_v \approx 2$.

## V. DISCUSSIONS

As can be seen above, although they show good agreement with the experimental values for the specific heat of a liquid, our results are slightly different from those obtained from the phonon theory of liquids due to the different methods used to calculate the energy of diffusion. In our approach, since the virial theorem is not used, anharmonic effects are not included when calculating the energy of diffusion. Hence, in the classical case, we obtain the same results as those from phonon theory of liquids for $\alpha = 0$.

In Sec. II, when calculating the energy of diffusion $E_d = K_d + P_d$, we neglected the potential parts of the energy of diffusing atoms, $P_d$, since $P_d \ll P_s(\omega > \omega_F)$ [13]. In our approach, if $P_d$ is known, these can be readily incorporated into the theory, as we have determined the number of diffusing atoms. $P_d$ can be introduced as a small parameter in the partition function, and with the aid of the calculated $N_d$, we can obtain $E_d$ in a similar way as for a gas.

The experimental heat capacity of liquids decreases as the temperature increases at a constant pressure, and rises as the pressure increases at a constant temperature. As can be seen from Eqs. (15) and (18), the theory explains their temperature dependence but does not account for their dependence on pressure. The pressure dependence enters the theory indirectly, through the use of experimental viscosity data. This leads to the important question of whether it is possible to model liquid heat capacity as a function of pressure and temperature without experimental viscosity data. In future studies, by integrating pressure dependence into our theoretical model, we plan to model liquid heat capacity without the need for viscosity.

## VI. CONCLUSIONS

By taking into account the numbers of oscillating and diffusing atoms, we have proposed a new method as a possible way to calculate the energy of diffusion without using the virial theorem. We calculated the energy of a subcritical liquid for both the quantum and the classical cases, and compared the results with the experimental specific heat of mercury, finding good agreement. Finally, we obtained an expression for the energy of a supercritical system. In further work, we aim to integrate pressure dependence into our theoretical model, to enable the prediction of liquid heat capacity without considering experimental viscosity data.